\documentclass[12pt]{article}
\usepackage{a4}
\usepackage[dvips]{graphicx}
\begin{document}

\centerline{ \bf Dirac quantization of the massless Thirring
model:}\centerline{ \bf  energy-momentum tensor anomaly}
\vspace*{1.0cm}
\renewcommand{\thefootnote}{\fnsymbol{footnote}}
\centerline{Sergei Kryukov\footnote{e-mail
sergei.kryukov@uleth.ca}} \vspace*{0.5cm}

\centerline  { Department of Physics, University of Lethbridge,}
\centerline { 4401 University Drive, Lethbridge, Alberta,\
T1K~~3M4, Canada.} \vspace*{1.5cm}

\centerline  {\bf Abstract} \vspace*{0.25cm}

 The Dirac method of quantizing Hamiltonian systems with
constraints is applied to the massless Thirring model. We solve the
quantum Hamiltonian equation for the energy-momentum
tensor and obtain a violation of the classical conservation law. 
A previously noticed
problem with the equal-time anticommutators can be fixed using this  Hamiltonian method.
\vspace*{0.25cm}

\noindent Classification codes: PACS 02.10.-v,03.65-w \hfill\break

\vspace*{1.5cm} \centerline {\bf 1. Introduction} \vspace*{0.25cm}

Since the late 1960's an extensive literature has evolved on the
massless Thirring model. We would like to fill a gap, however.

The Thirring model is a typical system with constraints and so we
quantize it here according to Dirac's special Hamiltonian
formalism. Unlike most authors, however, we don't use
 solutions of the normal-ordered Lagrange equations [1-4] in our
treatment, since such solutions are not part of the
Dirac-Hamiltonian formalism [5]. We only consider operators,
commutators, and normal ordering in initial time-like elements.
Only Hamiltonian language is used and the quantum Hamiltonian
equations of motion are solved. We  don't mix
elements of the Lagrange and Hamiltonian formalisms.

This method was successfully tested in the quantum integrable sine-Gordon, Zhiber-Shabat,
nonlinear Schrodinger, Korteweg-de Vries and modified Korteweg-de Vries  systems [6,7].
Quantum commutative integrals of motion for these models were constructed as  solutions of the quantum Hamiltonian equations.
 This is the unique known nonperturbative way of constructing quantum 
integrals  of motion, at the present time [6].
In addition,  using this method in the massless Thirring model helps to reveal a  new infinite dimensional symmetry in the sine-Gordon theory,
explaining the power law behavior of its correlation functions [7].

Furthermore using this Hamiltonian approach in the massless Thirring model we avoid a  problem with equal-time anticommutators. In fact we exchange this problem in its hidden form (in the equal-time  anticommutator) for an  obvious  anomaly. 
 Indeed the difference between the
standard equal-time anticommutator from the canonical one is of order $g^{2}$ for ($g \ll 1$). We fix this problem in the anticommutator, but get instead  a correction  in the same order 
of $g$ in another place: in the degree of 
non-analyticity of the
operator $T_{++}$. That  is the main result of this article.
It is important because this anomaly breaks the well known conformal symmetry of the massless Thirring model.

In Sec. 2 we  consider the  naive solution of this model in a modern, adapted version [3] and  explain the unsolved problem with this approach.
In Sec. 3 we use the classical Hamiltonian formalism for systems
with constraints on the massless Thirring model. We solve the
simple constraints and calculate their Dirac brackets. More
complicated constraints have complicated Dirac brackets and so we
work with them in a special way: very useful operators $P_{\pm}$
are introduced. In Sec. 4 we establish an important property of
initial functions, and so we use these functions in all operators
of our theory. In Sec. 5 we calculate a quantum anomaly of the
components of the energy-momentum tensor, a quantum correction to
the classical conservation law. Sec. 6 is devoted to concluding
remarks and a brief discussion of the connection of this work with
the Thirring/sine-Gordon equivalence.

\vspace*{1.0cm} \centerline  
{\bf 2. Problem with equal-time anticommutator } \vspace*{0.25cm}

Let us recall aspects of the  massless Thirring model in the modern conformal field theory approach.
We have the   vertex operator [3]
$$        
V_{m,n}(z,\bar z)=:\exp2i\left(\frac{\beta_{+}m+\beta_{-}n}{2}\epsilon(z)+
\frac{\beta_{+}m-\beta_{-}n}{2}\bar{\epsilon} (\bar z)\right):\eqno(2.1)
$$
so that $\epsilon(z,\bar z)$
satisfies equation  $\partial_{z}\partial_{\bar z} \epsilon(z,\bar z)=0  $, where 
$\epsilon(z,\bar z)=
\epsilon(z)+\bar {\epsilon}(\bar z)$. Components of the energy-momentum tensor are given by
$T(z)\sim :(\epsilon(z))^{2}:$ and ~~
$\bar{T}(\bar z)\sim :(\bar {\epsilon}(\bar{z}))^{2}:$.
In the standard approach these are holomorphic and antiholomorphic quantities.
Conformal dimensions have the form
$$
(\Delta,\bar \Delta)=(\beta_{+}m+\beta_{-}n)^{2},(\beta_{+}m-\beta_{-}n)^{2}),~~~ \beta_{+}\beta_{-}=\frac{1}{2},\eqno(2.2)
$$
$$
\beta_{+}=\left(\frac{1+g}{2(1-g)}\right)^{\frac{1}{2}},~~~
\beta_{-}=\left(\frac{1-g}{2(1+g)}\right)^{\frac{1}{2}}.\eqno (2.3)
$$ 
We can consider the obvious equations
$$
\partial_{\bar z}V_{m,n}(z,\bar z)=
:i(\beta_{+}m-\beta_{-}n)\partial_{\bar z}\bar {\epsilon} (\bar z))V_{m,n}:,
$$
$$
\partial_{z}V_{m,n}(z,\bar z)=:i(\beta_{+}m+\beta_{-}n)
\partial_{z}\epsilon (\bar z))V_{m,n}:.\eqno(2.4)
$$
After identifying certain vertex operators with fields of theory 
$$
\psi_{1}=V_{-\frac{1}{2}-\frac{1}{2}},~~~
\psi_{2}=V_{\frac{1}{2}-\frac{1}{2}}~~~,
\psi_{1}^{+}=V_{\frac{1}{2}\frac{1}{2}},~~~
\psi_{2}^{+}=V_{-\frac{1}{2}\frac{1}{2}},\eqno(2.5)
$$
we have equations for massless Thirring model
$$
\partial_{\bar z}\psi_{1}=g:\bar {J}\psi_{1}:,~~~
\partial_{ z}\psi_{2}=g:J\psi_{1}:.\eqno(2.6)
$$
These solutions for $\psi_{1},\psi_{2}$
satisfy normal ordered Lagrange equations.

But for a quantum solution we also must demand the correct equal-time anticommutator 
$$
[\psi_{1}(x,t),\psi_{1}^{+}(y,t)]_{+}=i\delta(x-y).\eqno(2.7)
$$
This anticommutator is a fundamental property of the quantum theory.
In the considered  case above the operator algebra of the 
solutions has the form
$$
\psi_{1}^{+}(z,\bar z)\psi_{1}(0,0)=z^{-\frac{1}{1-g^{2}}}\bar z ^{-\frac {g^{2}}{1-g^{2}}}.\eqno(2.8)
$$
In fact it is possible  to prove the equal-time property (which follows from the above) only for the $g=0$ case. And so this solution has a problem,  as indeed we will show  the quantum solution does.

Let us  consider one  example of an explanation in the literature [4]. The 
author considers the  solution 
$$
\psi(x)=\exp[-ib\gamma^{5}\widetilde {\phi}^{(-)}]
\exp[ia\phi^{(-)}(x)]\exp[ia\phi^{(+)}(x)]
\exp[-ib\gamma^{5}\widetilde {\phi}^{(+)}(x)]u \eqno(2.9)
$$
of the Thirring model,
where $u$ is a ``two component c-number quantity".
For the  equal-time anticommutator 
$$
<0|[\psi_{r}(x),\psi_{r}^{+}(y)]_{+}|0>=
u_{r}(x)u_{r}^{+}(y)\delta(x-y)(x-y)^{h-1}\eqno(2.10)
$$
is obtained.
Here $h=\frac{a^{2}+b^{2}}{2\pi}\geq 1$, $a$
and $b$ are constants in the theory.
By requiring
$$
u_{r}(x)u_{r}^{+}(y)\sim (x-y)^{-(h-1)},\eqno(2.11)
$$ 
the author gets the ordinary $\delta$ function result for the equal-time anticommutator. But we believe that  it is not possible to construct this ``c-number quantity" in a space of functions.
Indeed we have from (2.11) 
$$
\log u_{r}(x)+\log u^{+}_{r}(y)=(1-h)\log(x-y),\eqno(2.12)
$$
(for noncommutative operators 
$\hat a,\hat b$; we have  
 $\log (\hat a  \hat b) \ne \log \hat a +\log \hat b$).
If  $\partial_{x}\partial_{y}$ operator acts
on the left side, (2.12) we get 0;  on 
the right side we get
$\frac{(1-h)}{(x-y)^{2}}\ne 0$.

In the Hamiltonian approach, the canonical equal-time relation is postulated at the beginning  of the
calculation  and is preserved in time evolution, and
so  this problem disappears.

If we consider the massive Thirring model we get a similar problem with the  equal-time relation for fermions when $g \ne 0$ [8]
$$
[\psi(x,t),\psi^{+}(y,t)]_{+}=(x-y)^{\sigma(g^{2})}\delta(x-y)\ne\delta(x-y).\eqno(2.13)
$$
This problem can also  be fixed by the Hamiltonian method of quantization. We will consider the massive case in a separate work.

\vspace*{1.0cm} \centerline  {\bf 3. Classical Hamiltonian
formalism} \vspace*{0.25cm}

We use light cone coordinates with the notation
$$
x^{\pm}=\frac{1}{\sqrt{2}}(x^{0} \pm x^{1}),\eqno(3.1)
$$ and represent the $\gamma^{\pm}$-matrices by
$$
\gamma^{+}=
\sqrt{2}
\left ( \begin{array}{cc}
0&1\\
0&0\end {array} \right ),~~~
\gamma^{-}=
\sqrt{2}
\left ( \begin{array}{cc}
0&0\\
1&0\end {array} \right ).~~~\eqno(3.2)
$$
If we think of $x^{0}$ as the imaginary time $i\tau$, we work in
Euclidean space, and $x^{+}\sim z$,\ $x^{-}\sim \bar z$ are its
complex coordinates.

We start with the Lagrangian for the massless Thirring model in
light cone coordinates
$$
{\cal L}\ =\ \frac{i\sqrt{2}}{2}\left (\psi_2^+\partial_{+}\psi_2+
\psi_1^{+}\partial_{-}\psi_{1}-
\partial_{-}\psi_{1}^{+}\psi_{1}
-\partial_{+}\psi_{2}^{+}\psi_{2} \right )
-2g\,\psi_{1}^{+}\psi_{1}\psi_{2}^{+}\psi_{2}\ . \eqno(3.3)
$$
The canonical conjugates of the fields are the following:
$$
\pi_{\psi_{1}}=\frac{\partial L}{\partial (\partial_{-}\psi_{1})}
=-\frac{i\sqrt{2}}{2}\psi_{1}^{+},
~~~f^{1}_{1}=\frac{i\sqrt{2}}{2}\psi_{1}^{+}+\pi_{\psi_{1}},
$$
$$
\pi_{\psi_{1}^{+}}=\frac{\partial L}{\partial (\partial_{-}\psi_{1}^{+})}
=-\frac{i\sqrt{2}}{2}\psi_{1},
~~~f^{1}_{2}=\frac{i\sqrt{2}}{2}\psi_{1}+\pi_{\psi_{1}^{+}},\eqno(3.4)
$$
$$
\pi_{\psi_{2}}=\frac{\partial L}{\partial (\partial_{-} \psi_{2})}=0,
~~~f^{1}_{3}=\pi_{\psi_{2}},
$$
$$
\pi_{\psi_{2}^{+}}=\frac{\partial L}{\partial (\partial_{-}\psi_{2}^{+})}=0,
~~~f^{1}_{4}=\pi_{\psi_{2}^{+}}.
$$
We must have the canonical Poisson brackets (more exactly
Poisson-Berezin brackets for our anti-commutative variables) for
the fields and their conjugate fields:
$$
\{\psi_{1}(x);\pi_{\psi_{1}}(y) \}=
\delta(x-y),~~~
\{\psi_{1}(x);\pi_{\psi_{1}}(y) \}=
\delta(x-y),
$$
$$
\{\psi_{2}(x);\pi_{\psi_{2}}(y) \}
=\delta(x-y),~~~
\{\psi_{2}^{+}(x);\pi_{\psi_{2}^{+}}(y) \}
=\delta(x-y).\eqno(3.5)
$$
The expressions $f_{i}^{1}=0, (i=1,\ldots, 4)$ are the primary 
constraints. We have the first step Hamiltonian density,
$$
{\cal H}_{1}=-\frac{i\sqrt{2}}{2}\left
(\psi_{2}^{+}\partial_{+}\psi_{2}
-\partial_{+}\psi_{2}^{+}\psi_{2}\right )+
2g\psi_{1}^{+}\psi_{1}\psi_{2}^{+}\psi_{2}+ \lambda_{1}^{1}\left
(\frac{i\sqrt{2}}{2}\psi_{1}^{+}+\pi_{\psi_{1}}\right )+
$$
$$
\lambda_{2}^{1}\left
(\frac{i\sqrt{2}}{2}\psi_{1}+\pi_{\psi_{1}^{+}}\right )+
\lambda_{3}^{1}\pi_{\psi_{2}}+ \lambda_{4}^{1}\pi_{\psi_{2}^{+}}, \eqno(3.6)
$$
constructed in the usual way for systems with constraints [5]. We
must demand the conservation of constraints in time (a dot
indicates a derivative with respect to ``time'' $x_-$). From $\dot
f_{1}^{1}=\dot f_{1}^{2}=0$, we can obtain $\lambda
_{1}^{1},\lambda_{1}^{2}$, and from $\dot f_{3}^{1}=\dot
f_{4}^{1}=0$, we find new (secondary) constraints:
$f_{3}^{2},f_{4}^{2}$
$$
\dot f^{1}_{1}=\{H_{1};f^{1}_{1} \}=0;
~~~\lambda_{1}^{1}=\frac{2g}{i\sqrt{2}}\psi_{1}\psi_{2}^{+}\psi_{2},
$$
$$
\dot f^{1}_{2}=\{H_{1};f^{1}_{2} \}=0;
~~~\lambda_{2}^{1}=-\frac{2g}{i\sqrt{2}}\psi_{1}^{+}\psi_{2}^{+}\psi_{2},
$$
$$
\dot f^{1}_{3}=\{H_{1};f^{1}_{3} \}=0;
~~~f_{3}^{2}=i\sqrt{2}\partial_{+}\psi_{2}-2g\psi_{1}^{+}\psi_{1}\psi_{2}=0,\eqno(3.7)
$$
$$
\dot f^{1}_{4}=\{H_{1};f^{1}_{4} \}=0;
~~~f_{4}^{2}=i\sqrt{2}\partial_{+}\psi_{2}^{+}+2g\psi_{1}^{+}\psi_{1}\psi_{2}^{+}=0.
$$
Here we have introduced $H_{i}=\int h_{i}dz$. The constraints
$f_{3}^{2},f_{4}^{2}$ are part of the Lagrange equations but in
the Hamiltonian sense, they are only constraints of second class.
We must introduce the second step Hamiltonian density ${\cal
H}_{2}$:
$$
{\cal H}_{2}=-\frac{i\sqrt{2}}{2}\left
(\psi_{2}^{+}\partial_{+}\psi_{2}
-\partial_{+}\psi_{2}^{+}\psi_{2}\right )+
2g\psi_{1}^{+}\psi_{1}\psi_{2}^{+}\psi_{2}+
2g\psi_{1}\psi_{2}^{+}\psi_{2}\left
(\frac{i\sqrt{2}}{2}\psi_{1}^{+}+\pi_{\psi_{1}}\right )
$$
$$
-2g\psi_{1}^{+}\psi_{2}^{+}\psi_{2}
\left (\frac{i\sqrt{2}}{2}\psi_{1}+\pi_{\psi_{1}^{+}}\right )+
\lambda_{3}^{1}\pi_{\psi_{2}}+
\lambda_{4}^{1}\pi_{\psi_{2}^{+}}+ \eqno(3.8)
$$
$$
+\lambda_{3}^{2}
\left (i\sqrt{2}\partial_{+}\psi_{2}-2g\psi_{1}^{+}\psi_{1}\psi_{2}\right )+
\lambda_{4}^{2}
\left (i\sqrt{2}\partial_{+}\psi_{2}^{+}+2g\psi_{1}^{+}\psi_{1}\psi_{2}^{+}\right ).
$$
Demanding the conservation of the new constraints in time yields
$$
\dot f^{2}_{3}=\{H_{2};f^{2}_{3} \}=0;
~~~\lambda_{3}^{1},
$$
$$
\dot f^{2}_{4}=\{H_{2};f^{2}_{4} \}=0;
~~~\lambda_{4}^{1},\eqno(3.9)
$$
$$
\dot f^{1}_{3}=\{H_{2};f^{1}_{3} \}=0;
~~~\lambda_{3}^{2}=0,
$$
$$
\dot f^{1}_{4}=\{H_{2};f^{1}_{4} \}=0;
~~~\lambda_{4}^{2}=0.
$$
From $\dot f_{3}^{2}= \dot f_{4}^{2}=0$  we can determine
$\lambda_{3}^{1},\lambda_{4}^{1}$, but their forms are not
important for this work. Similarly, $\dot f_{3}^{1}=\dot
f_{4}^{1}=0$ determine  $\lambda_{3}^{2},\lambda_{4}^{2}$. Thus
all the constants in the ${\cal H}_{2}$ Hamiltonian density (3.8)
are determined and we have no new constraints.

It is very useful to resolve the constraints $f_{1}^{1}=
f_{2}^{1}=0$ and so we must calculate Dirac brackets. Using the
notation ($f_{1}=f_{1}^{1};f_{2}=f_{2}^{1}, \alpha, \beta=1,2$)
the expressions for the Dirac brackets are [5]
$$
\{\psi_{1},\psi_{1}^{+}\}_{\bf Dirac}=
\{\psi_{1},\psi_{1}^{+}\}-
\sum_{\alpha, \beta}\{\psi_{1},f_{\alpha}\}
\{f_{\alpha},f_{\beta}\}^{-1}\{f_{\alpha},\psi_{1}^{+}\},
$$
$$
\{\psi_{1}(x),\psi_{1}^{+}(y)\}_{\bf Dirac}=\frac{i}{\sqrt{2}}\delta(x-y),\eqno(3.10)
$$
$$
\{\psi_{1}(x),\psi_{1}(y)\}_{\bf Dirac}=
\{\psi_{1}(x)^{+},\psi_{1}^{+}(y)\}_{\bf Dirac}=0.
$$

\noindent In our case, the matrix of the constraints
$\{f_{\alpha}f_{\beta}\}$ is not degenerate, and so has an
inverse. We used $\delta^{-1}(x-y)=\delta (x-y)$. In our theory
the physical anti-commutative variables are
$\psi_{1},\psi_{1}^{+}$ only, and the anti-commutative variables
$\psi_{2},\psi_{2}^{+}$ are dependent. We don't resolve the
$f_{3}^{1}$, $f_{4}^{1}$, $f_{3}^{2}$, $f_{4}^{2}$ constraints,
and so we will use the ordinary Poisson (3.5) brackets between the
fields
 $\psi_{2},\psi_{2^{+}}$ and $\pi_{\psi_{2}},\pi_{\psi_{2}^{+}}$, but
 then impose the constraints.

After resolving the constraints $f_{1}^{1}=f_{2}^{1}=0$, the
Hamiltonian density ${\cal H}_{2}$ has the form:
$$
{\cal H}_{2}=-\frac{i\sqrt{2}}{2}\left
(\psi_{2}^{+}\partial_{+}\psi_{2}
-\partial_{+}\psi_{2}^{+}\psi_{2} \right )+
2g\psi_{1}^{+}\psi_{1}\psi_{2}^{+}\psi_{2}+
\lambda_{3}^{1}\pi_{\psi_{2}}+ \lambda_{4}^{1}\pi_{\psi_{2}^{+}}.\eqno(3.11)
$$
In our research, we are only interested in functionals
$I[\psi_{1},\psi_{1}^{+}]$. The important part of ${\cal H}_{2}$
is therefore ${\cal H}_{2}^{'}=
2g\psi_{1}^{+}\psi_{1}\psi_{2}^{+}\psi_{2},$ since $\{{\cal
H}_{2},I[\psi_{1},\psi_{1}^{+}]\}=\{{\cal
H}_{2}^{'},I[\psi_{1},\psi_{1}^{+}]\}$.

An important remark can now be made. The constraints $f_{3}^{2},
f_{4}^{2}$ must be ``hamiltonized''. We must introduce fields
$P_{\pm}$ as integrals over certain densities:
$$
P_{+}=\frac{2g}{i\sqrt{2}}\int\psi_{1}^{+}\psi_{1}\psi_{2}\pi_{\psi_{2}}\,dz\
,~~~ P_{-}=-\frac{2g}{i\sqrt{2}}\int\psi_{1}^{+}
\psi_{1}\psi_{2}^{+}\pi_{\psi_{2}^{+}}\,dz\ .\eqno(3.12)
$$
Here $\pi_{\psi_{2}}$ and $\pi_{\psi_{2}^{+}}$ are our constraints
$f_{3}^{1}$ and $f_{4}^{1}$. The operators $P_{\pm}$ do not
vanish, however, because their action is defined so that the
constraints are imposed only after calculating the Poisson
brackets.

After quantization, the operators $\hat P_{\pm}$ help to remove
singularities in the theory. We calculate
$$
\partial_{+}\psi_{2}=\{P_{+},\psi_{2}\}=
\frac{2g}{i\sqrt{2}}\{\int
\psi_{1}^{+}\psi_{1}\psi_{2}\pi_{\psi_{2}}dz,\psi_{2}\}
=\frac{2g}{i\sqrt{2}}\psi_{1}^{+}\psi_{1}\psi_{2}\,\eqno(3.13)
$$
using $\{\psi_{2}(x),\pi_{\psi_{2}(y)}\}=\delta(x-y)$, and obtain
the constraint $f_{3}^{2}$. Another important example of the
action of these operators is
$$
\partial_{+}(\psi_{2})\psi_{1}=\{P_{+},\psi_{2}\psi_{1}\}=
\frac{2g}{i\sqrt{2}}\{\int \psi_{1}^{+}\psi_{1}\psi_{2}\pi_{\psi_{2}}dz,\psi_{2}\psi_{1}\}=
$$
$$
=\frac{2g}{i\sqrt{2}}\psi_{1}^{+}\psi_{1}\psi_{2}\psi_{1}+
g\psi_{1}\psi_{2}\pi_{\psi_{2}}\psi_{2}=
\frac{2g}{i\sqrt{2}}\psi_{1}^{+}\psi_{1}\psi_{2}\psi_{1},\eqno(3.14)
$$
where we used
$$
\{\psi_{1}(x),\psi_{1}^{+}(y)\}_{\bf Dirac}=\frac{i}{\sqrt{2}}\delta(x-y),~~
\{\psi_{1}(x),\psi_{1}(y)\}_{\bf Dirac}=0,
$$
$$
\{\psi_{2}(x),\pi_{\psi_{2}}(y)\}=\delta(x-y),~~
\pi_{\psi_{2}}(x)=0. \eqno(3.15)
$$
While $\partial_{+}$ acts on $\psi_{2}$ only, $P_{+}$
acts on all fields $\psi_{2}$ and $\psi_{1}$, and imposing the
constraints removes the extra parts in the classical case. We will
consider the quantum analog of this example, and will find a
nontrivial action of this operator. Expression (3.14) is equal to
zero because $\psi_{1}^{2}=0$. We must remember here that
constraints $\pi_{\psi_{2}}=\pi_{\psi_{2}^{+}}  =0$ must be
imposed after calculating Poisson brackets (or commutators
(anti-commutators) in the quantum case).

\vspace*{1.0cm} \centerline {\bf 4. Quantization }
\vspace*{0.25cm}

Recall the solution of the quantum Hamilton equation
$$
\hat \psi(z,\bar z)\ =\ \exp\left(-i\bar z \hat H\right)\, \hat
\psi^{0}(z,\bar z)\, \exp \left(i\bar z\hat H\right).
\eqno(4.1)
$$
The superscript 0 indicates an initial quantum field, and
$\hat \psi(z,\bar z)$ denotes an operator solution. Let us
consider the operator Hamilton equation of motion:
$$
i\partial_{-}\hat \psi(z,\bar z )= [\hat H ,\hat \psi(z,\bar z)]\eqno(4.2)
$$
Inserting the solution gives
$$
i\partial_{-}\hat \psi(z,\bar z)\ =\ [\hat H,\hat\psi(z,\bar z)]
+i\exp\left(-i\bar z\hat H\right)\,
\partial_{-}\hat \psi^{0}(z,\bar z)\,
\exp\left(i\bar z\hat H\right)\eqno(4.3)
$$
For consistency then, we must demand that the initial operator
obey
$$
\partial_{-}\hat \psi^{0}(z,\bar z )\ =\ 0.\eqno (4.4)
$$
This is a very useful property (that of analytical functions) for
quantization.

It is a simple exercise to check that the solution above satisfies
fermionic properties too. For example, $\hat \psi^{2}(z,\bar z)=0$
(this follows from the fermionic properties of the initial quantum
fields).
It is easy to check property in Hamiltonian approach 
$[\hat {\psi}(z,\bar z),\hat {\psi}^{+}(z',\bar z)]_{+}=i\delta(z-z')$
(we have postulated for initial operators 
$[\hat {\psi}^{0}(z,\bar z),\hat {\psi}^{0+}(z',\bar z)]_{+}=i\delta(z-z'))$.
And so we don't have the problems like in naive approach.

 We will use the initial fields in all operators ($\hat
H, \hat P_{\pm}$) and will drop the ``0'' and $\bar z$ in the
notation $\hat \psi^{0}(z,\bar z)$. The Poisson brackets above
(for the classical initial fields) can be quantized in the usual
way, and we obtain the standard singular parts of operator product
expansions:
$$
\hat \psi_{1} (z)\hat \psi_{1}^{+} (z') =\frac{1}{(z-z')}\ ,
$$
$$
\hat \psi_{2} (z)\hat \pi_{\psi_{2}}(z') =\frac{1}{(z-z')}\ ,~~~
\hat \psi_{2}^{+} (z)\hat \pi_{\psi_{2}^{+}}(z')
=\frac{1}{(z-z')}.\eqno(4.5)
$$

\noindent Here we used the standard expansions of the analytical
fields:
$$
\psi_{1}(\xi)=\sum_{n}\psi^{n}_{1}(\xi-\alpha)^{-n-\frac{1}{2}},~~~
\psi_{1}^{+}(\xi)=\sum_{n}\psi^{+~n}_{1}(\xi-\alpha)^{-n-\frac{1}{2}}
,
$$
$$
\psi_{2}(\xi)=\sum_{n}\psi^{n}_{2}(\xi-\alpha)^{-n-\frac{1}{2}},~~~
\psi_{2}^{+}(\xi)=\sum_{n}\psi^{+~n}_{2}(\xi-\alpha)^{-n-\frac{1}{2}}\eqno(4.6)
,
$$
$$
\pi_{\psi_{2}}(\xi)=\sum_{n}\pi_{\psi_{2}}^{n}
(\xi-\alpha)^{-n-\frac{1}{2}},~~~
\pi_{\psi_{2}^{+}}(\xi)=
\sum_{n}\pi_{\psi_{2}^{+}}^{n}(\xi-\alpha)^{-n-\frac{1}{2}}
 ,
$$

\noindent where $\alpha$ is the center of the expansion. We have
also introduced some redetermine all of the fields multiply by
unimportant constants.

Poisson brackets for the complicated operators $\hat H,\hat
P_{\pm}$ must also be quantized. The integration contours for both
parts of the commutator ($\hat Q=\hat H,\hat P_{\pm}$):
$$
[\hat Q ,\hat A(\xi') ]=\hat Q \hat A(\xi') -\hat A(\xi') \hat Q ,~~~ \hat Q=\int
\hat q(\xi) d\xi.\eqno(4.7)
$$
need to be determined. Recall also that we are interested in only
a certain part of the Hamiltonian, $H^{'}_{2}$. Our quantum
operators are
$$
\hat H_{2}^{'}= 2g\int:\hat \psi_{1}^{+}\hat \psi_{1}\hat
\psi_{2}^{+}\hat \psi_{2}:d\xi\ ,
$$
$$
\hat P_{+}=\frac{2g}{i\sqrt{2}}\int:\hat \psi_{1}^{+}\hat \psi_{1}
\hat \psi_{2}\hat \pi_{\psi_{2}}:dx\ ,~~~ \hat
P_{-}=-\frac{2g}{i\sqrt{2}}\int:\hat \psi_{1}^{+}\hat \psi_{1}
\hat \psi_{2}^{+}\hat \pi_{\psi_{2}^{+}}:d\xi\ ,\eqno(4.8)
$$

\noindent where :: denotes normal ordering at the initial time.
Let us consider the contour integration in our commutators. In the
first part of the commutator, we choose the contour closing above
the point $\xi'$ (see Fig.1).
\begin{figure}[h]
\centering
\includegraphics {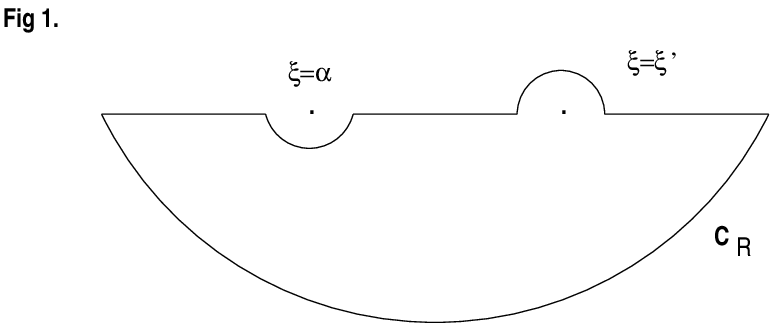}
\centerline {Fig.1 Contour of integration for commutator.}
\end{figure}

\noindent In the second part, we choose the contour closing below
the same point $\xi'$ (see Fig.2).
\begin{figure}[h]
\centering
\includegraphics {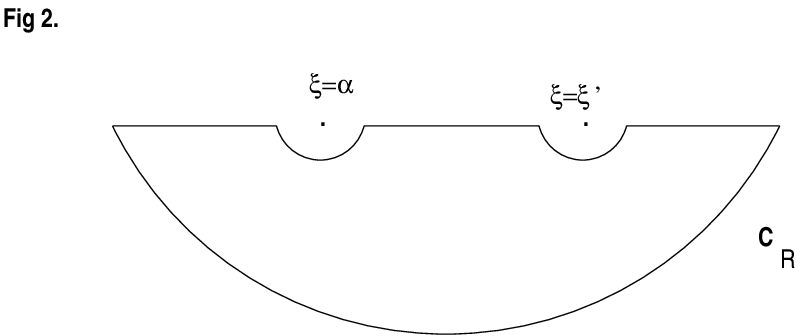}
\centerline{\small Fig. 2 Contour of the integration  for the
commutator.}
\end{figure}

Letting the radius $R$ of the semicircles go to $\infty$, we have
$$
\int_{C_{R\to \infty}}\hat {\cal H}_{2}^{'}(\xi) \hat A(\xi')
d\xi=0\ ,~~~ \int_{C_{R\to \infty}}\hat p_{\pm}(\xi) \hat A
(\xi')d\xi=0\ ,\eqno(4.9)
$$
using the asymptotic behavior of the fundamental fields $\hat
\psi_{1},\hat \psi_{1}^{+}\,\sim\,\frac{1}{R}$. So for our
operators $\hat H^{'}_{2}, \hat  P_{\pm}$, (using analytic
property (4.4)):
$$
[\hat Q ,\hat A(\xi') ]=\oint_{\xi'}\hat q (\xi) \hat A (\xi')
d\xi ,\eqno(4.10)
$$
where the notation indicates closed-contour integration around the
point $\xi=\xi'$.

Let us consider the action of the quantum operators $\hat P_{\pm}$
(4.8). In our case we must calculate expressions like $\partial
_{+}(\hat \psi_{2})\hat \psi_{1}^{+}$. The naive way (without
introducing $\hat P_{\pm}$ operators) is would be
$$
\lim_{z\to z'}\partial _{+}(\hat \psi_{2})(z)\hat
\psi_{1}^{+}(z')= \frac{2g}{i\sqrt{2}} \lim_{z\to z'}(\hat
\psi_{1}^{+}\hat \psi_{1}\hat \psi_{2})(z)\hat \psi_{1}^{+}(z').\eqno(4.11)
$$
We find a singularity for $ z \to z'$, arising from the product
$\hat \psi_{1}(z)\hat \psi_{1}^{+}(z')\sim \frac{1}{z-z'}$, so
that
$$
\lim_{z\to z'}(\hat \psi_{1}^{+}\hat \psi_{1}\hat \psi_{2})(z)\hat
\psi_{1}^{+}(z')=\infty .\eqno(4.12)
$$
If we use the action of the quantum operators $\hat P_{\pm}$, we
instead obtain
$$
\partial _{+}(\hat \psi_{2})\hat \psi_{1}^{+}=igh\partial_{+}
(\hat \psi_{1}^{+} \hat \psi_{2}).\eqno(4.13)
$$
The singularity has been removed, but there is a quantum
correction to the classical result ($h$ is Planck's constant).
This is the idea behind the introduction of these operators $\hat
P_{\pm}$. Using the quantum Hamiltonian $\hat H$, when we
calculate $[\hat H,\hat I]$ we find singularities. The analytic
property of initial fields (4.4), however, with the choice of the
contour of the integration described above help to remove those
singularities. Incidentally, this way (in $[\hat H,\hat I]$) of
removing the singularity gives the correct commutative integrals
of motion (elements of Hamiltonian formalism too) for quantum
sine-Gordon theory [7].

In this article we will solve the quantum Hamilton equation
$$
i\partial_{-}\hat A\ =\ [\hat H ,\hat A ] ,\eqno(4.14)
$$
for the ``++'' component of the energy-momentum tensor. That is,
we put $\hat A =\hat T_{++}=\hat T $, with
$$
\hat T=:\hat \psi_{1} \partial_{+}\hat \psi_{1}^{+}:+:\hat \psi_{1}^{+}
\partial_{+}\hat \psi_{1}: ,
\eqno(4.15)
$$
where  $\hat \psi_{1},\hat \psi_{1}^{+}$ and $\hat T$ are initial
functions. We must calculate $[\hat H ,\hat T]$.

\vspace*{1.0cm} \centerline {\bf  5. Calculation of anomaly  }
\vspace*{0.25cm}

Let us consider
$$
\hat i_{2}= :\hat \psi_{1} \partial_{+}\hat \psi_{1}^{+} :\ .\eqno(5.1)
$$
After a simple calculation we obtain
$$
[\hat H ,\hat i_{2}]=\eqno(5.2)
$$
$$
=2gi(-):\hat \psi_{1} \hat \psi_{2}^{+} \hat \psi_{2}
\partial_{+} \hat \psi_{1} :+
2gi(-):\partial_{+}(\hat \psi_{1}^{+} \hat \psi_{2}^{+}\hat
\psi_{2} ) \hat \psi_{1}^{+}: +gih\partial^{2}_{+}(\hat
\psi_{2}^{+} \hat \psi_{2})(-) .
$$
Similarly, for
$$
\hat i_{2}^{'}= -:\hat \psi_{1}^{+}\partial_{+}\hat \psi_{1}: ,\eqno(5.3)
$$
we find
$$
[\hat H ,\hat i_{2}{'}]=\eqno(5.4)
$$
$$
=2gi(-):\hat \psi_{1} \hat   \psi_{2}^{+}
\hat \psi_{2} \partial_{+} \hat \psi_{1}:+
2gi(-):\partial_{+}(\hat \psi_{1}^{+} \hat \psi_{2}^{+}\hat
\psi_{2} ) \hat \psi_{1}^{+}: +gih\partial^{2}_{+}(\hat
\psi_{2}^{+} \hat \psi_{2})(-) .
$$
The result is
$$
[\hat H,\hat T]\ =\
4gi:\hat\psi_{1}\partial_{+}(\hat\psi_{2}^{+}\hat\psi_{2})\hat\psi_{1}^{+}:
.\eqno(5.5)
$$
Now we can calculate $ :\hat \psi_{1} \hat \psi_{1}^{+}
\partial_{+}(\hat \psi_{2} \hat \psi_{2}^{+}): $ using our
operators $\hat P_{\pm}$. We find
$$\eqno(5.6)
$$
\begin{eqnarray*} :\partial_{+}
\hat \psi_{2}^{+} \hat \psi_{2} \hat \psi_{1}^{+} \hat \psi_{1} :\
&=&\ (-igh):\partial_{+} (\hat \psi_{1} \hat \psi_{2}^{+} ) \hat
\psi_{2} \hat \psi_{1}^{+} :\, +\, (-igh):\partial_{+} (\hat
\psi_{1}^{+} \hat \psi_{2}^{+} )\hat \psi_{2} \hat \psi_{1} :\\ &
& +\,(-igh^{2})\partial^{2}_{+}\hat
\psi_{2}^{+} \hat \psi_{2}\ ,\\
 :\hat \psi_{2}^{+} \partial_{+} \hat \psi_{2} \hat
\psi_{1}^{+} \hat \psi_{1}:\ &=&\ (-igh):\partial_{+} (\hat
\psi_{1} \hat \psi_{2}) \hat \psi_{2}^{+}\hat \psi_{1}^{+}: \, +\,
(-igh):\partial_{+} (\hat \psi_{1}^{+}\hat \psi_{2})\hat
\psi_{2}^{+} \hat \psi_{1}:\\
& & +\,(-igh^{2})\partial^{2}_{+}\hat\psi_{2}\hat \psi_{2}^{+}.
\end{eqnarray*}
We therefore have
$$
:-\hat \psi_{1} \hat \psi_{1}^{+} \partial_{+}(\hat \psi_{2} \hat
\psi_{2}^{+}):\ =\ -igh(\partial^{2}_{+}\hat \psi_{2} \hat
\psi_{2}^{+}\,+\,\partial^{2}_{+}\hat \psi_{2}^{+}\hat \psi_{2}) \eqno(5.7)
$$
yielding
$$
[\hat H, \hat T ] \ =\ -2\mu^{2}(\partial ^{2}_{+}\hat \psi_{2}
\hat \psi_{2}^{+}\,+\, \partial^{2}_{+} \hat \psi_{2}^{+} \hat
\psi_{2})\ ,~~~ \ \ \mu=igh .\eqno(5.8)
$$

If $[\hat H,\hat T]$ is not zero, then we have a quantum anomaly,
$\partial_{-}\hat T \neq 0$. This is easily established:
$$
(1-\frac{\mu^{2}}{2})
(\partial ^{2}_{+}\hat \psi_{2}\hat\psi_{2}^{+}+\partial^{2}_{+}
\hat \psi_{2}^{+}
\hat \psi_{2})=2ig
(:\partial_{+} \hat \psi_{1}^{+}
\hat \psi_{1}\hat \psi_{2}\hat \psi_{2}^{+}:
+ :\hat \psi_{1}^{+}
\partial_{+} \hat \psi_{1} \hat \psi_{2} \hat \psi_{2}^{+}:) \neq 0.\eqno(5.9)
$$

The ``++'' component of the energy-momentum tensor (4.15) is
therefore not conserved in time. However, if we have $h=0$
(classical limit) or $g=0$ (free massless fermions), we do have
$$
[\hat H,\hat T]=0,\eqno(5.10)
$$
and so $\hat T$ is conserve for a quantum free massless fermion
theory.

In the usual  quantum case only one mode (momentum) is conserved,
and indeed we can make a simple transformation to obtain
$$
[\hat H ,\hat T ]\ =\ 2\mu^{2}\partial_{+}(\partial_{+}\hat
\psi_{2} \hat \psi_{2}^{+}+\partial_{+} \hat
\psi_{2}^{+}\hat\psi_{2}) .\eqno(5.11)
$$
If we introduce notation for the momentum operator $\hat I_{2}
=\int \hat T dz$, we have $[\hat H ,\hat I_{2}]=0$. If we want to
calculate $\hat T_{-~-}$ component of the energy-momentum tensor
we must consider ``+'' variable like the time, from the beginning
of the calculation.

\vspace*{1.0cm} \centerline {\bf  6. Conclusion  }
\vspace*{0.25cm}

An important conclusion can now be drawn. We can bosonize the
initial functions using
$$
\hat \psi_{1}(z)\ =\ :\exp\left(\hat \phi(z)\right):\ ,~~~~~ \hat
\psi_{1}^{+}(z)\ =\ :\exp\left(-\hat \phi(z)\right): ,\eqno(6.1)
$$
where $\hat \phi$ is the initial bosonic function (operator).
After a simple transformation we find
$$
\hat T\ =\  :\hat \psi_{1}\partial_{+} \hat \psi_{1}^{+}: + :\hat
\psi_{1}^{+}\partial_{+} \hat \psi_{1}:\ =\ -:(\partial_{+}\hat
\phi)^{2}: .\eqno (6.2)
$$
We see the equivalence between the ``++'' component of the
energy-momentum tensor for bosonic and Thirring fermionic theories
at the initial time. A similar calculation can be found in [9]. In
the quantum massless Thirring model we have $\partial_{-}\hat
T_{\bf Thirring}\neq 0 $, however.  So, notwithstanding the
equivalence of the initial operators $\hat T$, the equivalence
between the massless Thirring model and the massless free bosonic
field is lost, if the Hamiltonian formalism is used throughout.
Indeed let us consider the free massless bosonic theory. It has
the Hamiltonian $\hat H^{\bf free}_{\bf boson}=0$ in light cone
coordinates, and so $\hat T^{\bf free}_{\bf boson}$ (6.2) is
conserved in time, $\partial _{-}\hat T^{\bf free}_{\bf boson}=0$.

We see then, the violation of the zeroth-order approximation (in
the sense of [10])  of the sine-Gordon /Thirring model equivalence
[8,10]. Of course, the solution [1-4] can still sometimes be useful,
because the violation is very weak, $\sim g^{2}h^{2}$, for $g \ll 1$.

\vspace*{1.0cm} \centerline {\bf  Acknowledgements  }
\vspace*{0.25cm} This work was supported by a NATO Science
Fellowship, the University of Lethbridge, and by NSERC of Canada.
We thank  Mark Walton and Jian-Ge Zhou for discussions.

\vspace*{1.25cm} \centerline {\bf  References  } \vspace*{0.25cm}
\parindent=0pt

[1] B. Klaiber, in Lectures in Theoretical Physics Lectures
delivered at the Summer Institute for Theoretical Physics
University of Colorado, Boulder, 1967, ed A.Barut, W.Brittin,
Gordon and Breach, New York, 1968, v.X, part A, p. 141-176.

[2] G.F. Dell'Antonio, Y. Frishman, D. Zwanziger, Phys. Rev. D6
(1972) 988.

[3] A.B.Zamolodchikov, Al.B.Zamolodchikov, Conformal Field Theory and Critical Phenomena in Two Dimensional Systems, Soviet Scientific Reviews / sect A v.10  part 4  (1989) 368.

[4] N.Nakanishi, Progress of Theoretical Physics  57 n.2 (1977) 580.

[5] D.M. Gitman, I.B. Tyutin, Canonical quantization for fields
with constraints,  (Nauka, Moscow, 1986),
D.M.Gitman,I.V.Tyutin, Quantization of Fields with Constraints (Springer -Verlag, Berlin, 1990).

[6] M. Omote, M. Sakagami, R. Sasaki, I. Yamanaka, Phys. Rev. D v.35 n.8  (1987) 2423.

~~R. Sasaki, I. Yamanaka, Comm. Math. Phys.  108 (1987) 691.

~~R. Sasaki, I. Yamanaka, Adv. Stud. in Pure. Math., v.16  (1988) 271.

~~R. Sasaki, I. Yamanaka, in Essays in Honor of the 60-th
Birthday of Professor 
Y. Yamaguchi, edited by H Terazawa (World Scientific, Singapure, 1986).

[7] S. Kryukov, Special integrals of motion in quantum integrable system, Preprint.

[8] S. Mandelstam, Phys. Rev. D11 (1975) 3026.

[9] D.Ts. Stoyanov, L.K. Hadjiivanov, About Quantum Energy
Momentum Tensor for the Thirring Model, JINR preprint, P2-84-466,
Dubna 1984.

[10] S. Coleman, Phys. Rev. D11 (1975) 2088.

\end{document}